\documentclass[aps,twocolumn,showpacs,superscriptaddress,%
floatfix]{revtex4-1} 
\usepackage{psfrag}
\usepackage[usenames,dvipsnames,svgnames,table]{xcolor}
\usepackage{amssymb,amsmath,amsthm}
\usepackage{subfig}
\usepackage[dvips]{graphicx}
\usepackage{verbatim}
\usepackage{graphicx}
\usepackage{dcolumn}
\usepackage{bm}
\usepackage[absolute]{textpos}
\usepackage{hyperref}

\usepackage{subfig}
\usepackage{float}

\begin{document}
\widetext
\title{Spin-coating of moderately concentrated superparamagnetic colloids in different magnetic field configurations}
\author{Raheema Aslam}\email{E-mail: rmuhammad@alumni.unav.es}\affiliation{Universidad de Navarra, Complex Systems Group, Pamplona, Spain}
\author{Wenceslao Gonz\'{a}lez-Vi\~{n}as}\email{E-mail: wens@unav.es}\affiliation{Universidad de Navarra, Complex Systems Group, Pamplona, Spain}\affiliation{Universidad de Navarra, PHYSMED Group, Pamplona, Spain}

\begin{abstract}
Spin-coating technique is very fast, cheap, reproducible, simple and needs less material to fabricate films of particulate systems/colloids. Their thickness and uniformity may be controlled by means of external fields. We apply magnetic fields during the spin-coating of a moderately concentrated superparamagnetic colloid (made of silica coated magnetite particles). We study the influence of different magnetic field configurations (homogeneous and inhomogeneous) on the resulting spin-coated deposits and compare experimental results under various conditions. Superparamagnetic colloids behave as, non-Newtonian, magnetorheological fluids. Their viscosity vary significantly under applied magnetic fields. We measure and compare the effect of uniform and non-uniform magnetic fields on their relative effective viscosity, using the spin-coated deposits and a previously existing model for simple colloids. The mechanisms involved in the deposits formation under different experimental conditions are also discussed. In particular, we show that the magnetophoretic effect plays an important role in the spin-coating of magnetic colloids subjected to non-uniform magnetic fields. We characterize an effective magnetoviscosity in non-uniform magnetic fields that is largely influenced by the magnetophoretic effect that enhances the flow of the magnetic fluid.
\end{abstract}

\maketitle
\begin{textblock}{18}(1.5,15)
\tt\textcopyright\ 2017. This manuscript version is made available under the CC-BY-NC-ND 4.0 \href{http://creativecommons.org/licenses/by-nc-nd/4.0/}{license}
\end{textblock}
\begin{textblock}{18}(1.5,0.15)
\setlength{\fboxrule}{2pt}\colorlet{currentcolor}{.}
{\color{red}%
    \fbox{\color{currentcolor}\parbox{15cm}{Accepted for publication in Colloids and Surf. A\\[0.3mm]The final publication is available under{\bf DOI: \href{https://dx.doi.org/10.1016/j.colsurfa.2017.04.007}{10.1016/j.colsurfa.2017.04.007}}.\\[0.52mm]Please, cite as: R. Aslam \& W. Gonz\'alez--Vi\~nas. Colloids Surf. A 532, 530--534 (2017)}}}
\end{textblock}
\section{Introduction}
Spin-coating technique has been introduced in colloids to fabricate thin films and colloidal crystals since long \cite{Walker1922, Emslie1958, Meyerhofer1978, Rehg1992, BirnieIII1997, Acrivos1960, Wahal1993, Furukawa2005, Bornside1989, Hall1998, Fraysse1994, Ohara1989, Mihi2006, Sahu2009}. It is simple, reproducible and requires little material to get results in a very short time \cite{Wu2007_thesis}. The thickness and uniformity of spin-coated films which are important parameters for their technological applications, depend on experimental conditions including; the spin time and speed, the viscosity of fluids, the density and the evaporation rate of the fluids, concentration of the suspension and the substrate surface characteristics. The flow behavior as a function of the fluid properties in spin-coating process helps to recognize its strong performance in applications and has been investigated by many research groups through different models and experiments \cite{Emslie1958, Meyerhofer1978, BirnieIII1997, Higgins1986, Zhao2008, Arcos2008, Giuliani2010, Cregan2007, Cregan2013}. Spin-coating of nano-colloids has been thoroughly explored in the literature for almost a century \cite{Walker1922}. However, for colloids of bigger particles, external fields have been used recently for tunning their effect which is not yet fully understood \cite{Pichumani2011a, Bartlett2012, Pichumani2013, Aslam2016}.

The rheological effects of magnetic fields on magnetic fluids have been studied from mid-twentieth century \cite{Neuringer1964,McTague1969,Shliomis1972}. One of the most important result is that the viscosity of a magnetic fluid increases monotonously on applying a magnetic field, until it reaches a saturation value. 

Superparamagnetic particles of typical size ranging from 100~nm to a few micrometers are prepared by inserting magnetic nanoparticles in a matrix of non-magnetic material (polystyrene or silica). They show a quasi-zero remanent magnetization and high magnetic response \cite{Faraudo2013}. They have many interesting applications in the biomedical field, e.g. drug delivery; to deliver medicines to arteries and veins in the circulatory system of the body, due to their high magnetic response \cite{Corchero2009}. To use them for this kind of applications, superparamagnetic colloids are first functionalized to capture their specific targets. Then, magnetic gradient is applied for removing them from their targets through magnetophoresis \cite{yavuz2009}. Where, magnetophoresis defines the motion of magnetic particles in a magnetic gradient. An magnetic field results in the formation of superparamagnetic particles chains along the magnetic field which enhances the process of separation \cite{darras2016}. As the homogeneous magnetic fields can not induce a drift velocity in magnetic particles therefore a magnetic gradient is required for having magnetophoresis \cite{Faraudo2013}. 

The effect of homogeneous external magnetic fields on the behavior of diluted simple superparamagnetic \cite{Pichumani2011a} as well as hybrid ferromagnetic \cite{Aslam2016} colloids in spin-coating have been studied. They observed an increase in the solid deposits on the substrate with applied magnetic field as compared to the sparse structures deposits obtained without applying magnetic fields. Later, they measured the magnetoviscosity of colloids from the amount of spin-coated deposits obtained without and with homegeneous magnetic fields \cite{Pichumani2013,Aslam2016}, by generalizing a continuum model \cite{Cregan2007} to particulate systems. In summary, they proved that external uniform magnetic fields only affect the magnetoviscosity (an increase in the viscosity when a magnetic field is applied \cite{McTague1969}) of magnetic colloids, and that the generalized model holds for diluted magnetic colloids. 

In this article,  we use the same previously proved generalized model for colloids \cite{Pichumani2013,Aslam2016}, to measure the magnetoviscosity as a function of different magnetic field configurations in spin-coating. Here, moderately concentrated suspension of superparamagnetic particles in a volatile solvent is used. We measure the surface coverage of the dried deposits obtained without, with uniform and with non-uniform magnetic fields that are applied during the spin-coating of the suspension. Firstly, the experimental setup and methods are described. After, the previously existing models \cite{Cregan2007,Pichumani2013,Aslam2016} for simple fluids and colloids are highlighted. Finally, the results obtained through spin-coating at various experimental conditions are presented and discussed. Our results corroborate that spin-coating technique allows to study magnetorheology in a very short time using a uniform magnetic field, and that the increase in effective viscosity of the colloids is the only effect of the applied magnetic field \cite{Pichumani2013,Aslam2016}. However, when the magnetic field is not uniform, there are other relevant effects, namely magnetophoretic effect, which leads to a decrease of the effective viscosity.

\section{Experimental methods}
\begin{figure*}[!ht]
\centering
  \includegraphics[width=14cm]{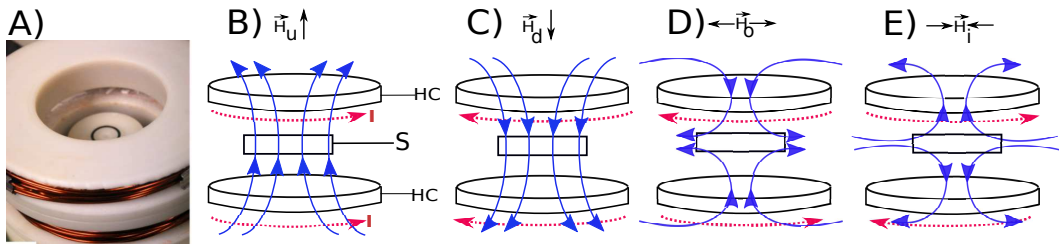}
  \caption{(A) Photograph of spin-coater including a pair of Helmholtz coils. Reproduced from \cite{Pichumani2013} with permission from the Royal Society of Chemistry. (B--D)~Sketches of experimental setup configurations with magnetic field lines (HC:~Helmholtz coils; S:~Substrate, I:~Electrical current). The substrate spins in the region of uniform magnetic field (B and C) and nonuniform magnetic fields (D and E).}
  \label{setupSC}
\end{figure*}

Experiments are performed in a customized commercial spin-coater (Laurell technologies, WS-650SZ-6NPP/LITE/OND) at rotation rate of 2000~rpm. The photograph of the spin-coater provided with Helmholtz coils is shown in Figure~\ref{setupSC}A. Sketches of the experimental setup evidencing applied magnetic field of various configurations obtained by a pair of Helmholtz coils which are placed in such a way that the substrate spins in the region of uniform axial (Figures~\ref{setupSC}B--C) and nonuniform fields (Figures~\ref{setupSC}D--E).
Different applied magnetic field configurations are generated by adjusting the orientation of a fixed electrical current (5~A) in these coils with an external power supply. Using this experimental setup we obtain four different configurations of the magnetic field which are detailed in the table \ref{MGF}. 
\begin{table}[!ht]
\centering
\begin{tabular}{ | l | l | l | p{5cm} |}
\hline
Configuration & Symbol & Magnetic field ($|$H$|$) [kA/m]\\ 
\hline
Zero-field & 0 & 0 \\ 
\hline
Up & H$_{u}$ & 10.145 $\pm$ 0.005 \\ 
\hline
Down & H$_{d}$ & 10.915 $\pm$ 0.005\\
\hline
Outward & H$_{o}$ & 0.61 $\pm$ 0.26\\  
\hline
Inward & H$_{i}$ & 0.99 $\pm$ 0.16\\ 
\hline
\end{tabular}
\caption{Various magnetic field configurations, their symbols and magnitudes, used for performing spin-coating experiments.\label{MGF}}
\end{table}

Superparamagnetic particles obtained from microParticles GmbH, Germany, that consist of silica coated magnetite of diameter 1.51$\pm$0.05~$\mu$m (density~=~1.6 -- 1.8~g/cm$^3$), are used for all experiments. Their magnetic properties can be found in the Electronic Supplementary Information of \cite{Pichumani2013}. The particles are dried overnight and then weighted for preparing suspension of concentration of 12.82~\%(v/v) in Methyl Ethyl Ketone (MEK). The estimated dynamic viscosity of the colloid (at zero-field) is 55\% bigger than the one of the solvent alone \cite{Guth1936}. The suspension is ultrasonicated for fifteen minutes before each experiment. The spin-coater is operated at 2000~rpm and the magnetic field $H$ is applied. Then, 100~$\mu$l of suspension is pipetted onto the spinning substrate. Once the spun suspension is dried, the field is turned off. Images of the dried deposits on the substrates are taken with a white light reflection microscope at 2~mm intervals from the center of spinning. After, we analyzed one micrograph for each set ($H$, $r$), where $r$ is the distance from the center of spinning. 

For all experiments, glass substrates of an approximate size of 38~x~25~x~1~mm$^3$ are used. First, they are cleaned in soft basic piranha for forty minutes at 67$^\circ$C (piranha is a mixture of ultra-pure water/ammonia/hydrogen peroxide with the ratio of 5:3:1 in volume). Next, they are rinsed with ultra-pure water. Finally, they are dried by a filtered air blow. 

\begin{figure}[!ht]
 \centering
 \includegraphics[width=8.5cm]{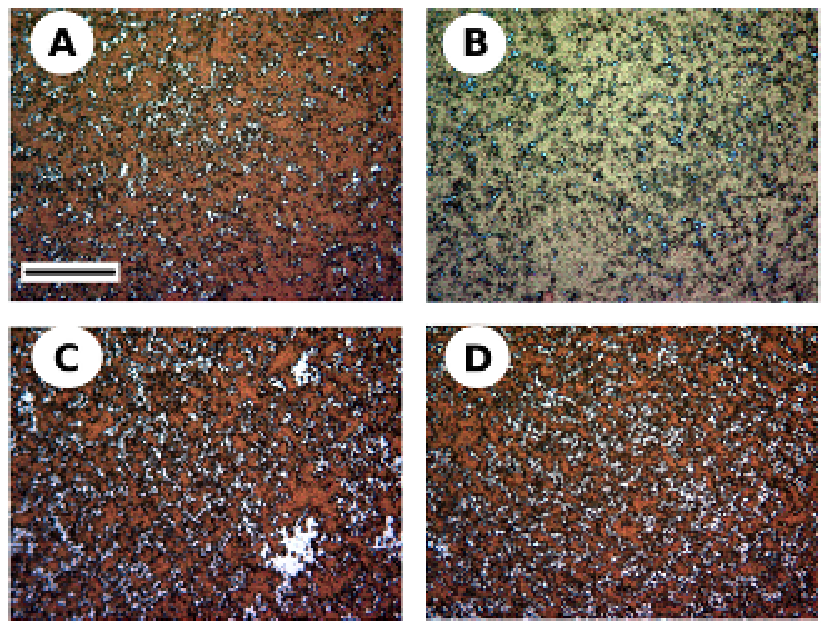}
\caption{Images of the dried spin-coated deposits on the substrate at 6~mm distance from the center of spinning. A) From the experiment performed at zero magnetic field. (B--D) From the experiments carried out at different applied magnetic fields; (B): H$_{u}$, (C) :H$_{o}$ and (D):~H$_{i}$. The corresponding magnetic fields values are given in the table \ref{MGF}. For all the cases the experimental conditions are: spinning speed of 2000~rpm, concentration of the suspension is 12.82~\%(v/v), and the diameters of silica coated magnetite particle are 1.51$\pm$0.05~$\mu$m. The scale bar is 164~$\mu$m}
  \label{deposits}
\end{figure}

\begin{figure}[!ht]
  \centering
  \includegraphics[width=8.5cm]{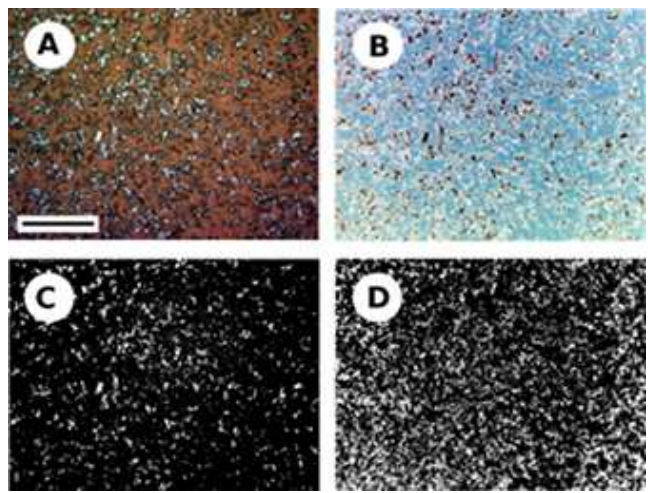}
  \caption{A) Micrograph of the spin-coated deposits from the experiment performed without magnetic field at 6~mm distance from the center of spinning with spinning rate of 2000~rpm, concentration of the suspension is 12.82~\%(v/v) in MEK, and the diameters of superparamagnetic particles are 1.51~$\mu$m. The scale bar is 164~$\mu$m. B) Color inverted image of (A). C,D) Relevant channels of RGB decomposition obtained by using home-made routines in Octave used for further analysis to obtain CEH.}
  \label{deposits_analysis}
\end{figure}

Figure~\ref{deposits} shows typical micrographs of spin-coated colloidal deposits obtained from the experiments performed at 2000~rpm, without magnetic field (Figure~\ref{deposits}A) and with applied magnetic field of various configurations (Figure~\ref{deposits}B--D) that are sketched in Figure~\ref{setupSC}B--E, respectively. As in Figure~\ref{deposits}, we see that the deposits from all the experiments are mainly mixtures of mono and bilayers. We identify the regions with monolayer, bilayer (and occasionally empty space) by using their wavelength dependence for Bragg reflected light \cite{Schoepe2006}. As color information is useful, for each micrograph (e.g. Figure~\ref{deposits_analysis}A), we invert the color field and adjust brightness and contrast (e.g. Figure~\ref{deposits_analysis}B), and perform RGB decomposition (where the relevant channels to our experiment can be seen, e.g., in Figure~\ref{deposits_analysis}C--D). Finally, we threshold the images, to characterize the amount of deposit using home-made routines in Octave, and use it for further analysis, in terms of occupation factor $\varepsilon_i^2$ (area occupied by particles in a specific layer $i$ divided by the substrate area).

\section{Theoretical background}
\subsection{Model of simple fluids.}

A model \cite{Cregan2007} by Cregan {\it et al.}, that depends on the lubrication approximation and uses the matched asymptotic expansions technique, provides a much better solution than the one obtained by Meyerhofer \cite{Meyerhofer1978} for spin-coating of simple fluids with evaporation. Among its assumptions the most important include the following; it considers the evaporation rate constant, the system is continuous (i.e. not particulate) and the fluid is Newtonian. The thickness (or height) of the final spin-coated film of Cregan's model depends on the parameters of the suspension and on the spin-coating procedure. The resultant equation giving the final deposited film thickness (h$\displaystyle_{\infty}^{(s)}$) by this model is:

\begin{equation}
h\displaystyle_{\infty}^{(s)} = \frac{C}{1-C}\left(\frac{3}{2}\nu E\right)^{\frac{1}{3}}\omega^{-\frac{2}{3}},\label{eqcregan-bis}
\end{equation}
where, $C$ is the volume fraction, $\omega$ is the rotation rate, $\nu$ is the kinematic viscosity of the solvent, and $E$ is the evaporation rate of solvent. 

This model holds also for colloidal systems \cite{Pichumani2013, Aslam2016}, once the appropriate adaptations have been taken.
 
\subsection{Colloids model.}

In a colloidal system, the effective local amount of matter deposited on the substrate depends on the geometry of deposited structure and is defined as the compact equivalent height (CEH), which is the thickness of a continuous homogeneous layer whose volume is the same as all the particles deposited. In general, it is obtained from the product of atomic packing fraction and the Voronoi cell \cite{Voronoi1908} volume. Shortly, for a colloidal or particulate system, $h\displaystyle_{\infty}^{(s)}$ of eq.~\ref{eqcregan-bis} for simple fluids, has to be replaced by CEH. Hence, eq.~\ref{eqcregan-bis} for the final colloidal film thickness becomes:
\begin{equation}
\mbox{CEH}= \frac{C}{1-C}\left(\frac{3}{2}\nu E\right)^{\frac{1}{3}}\omega^{-\frac{2}{3}},\label{eqcregan-ceh}
\end{equation}
where in our experiments $C=0.1282$ (v/v) and $\omega=2000$ rpm.

\begin{figure}[!ht]
\centering
\includegraphics[width=8cm]{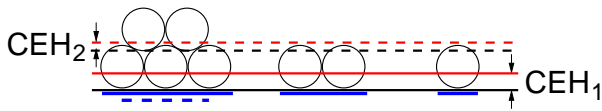}
\caption{Sketch for the construction of the compact equivalent height in colloidal deposits. $\mbox{CEH}_{1}$ is the contribution of the first layer particles (solid lines) and $\mbox{CEH}_{2}$ is the contribution of the second layer of the deposit (dashed lines).}
\label{cehsketch}
\end{figure}

For each layer $i$ of a colloidal deposit with a particular structure we can write:  

\begin{equation}
\mbox{CEH}_i = K\cdot\varepsilon_i^2,\label{ceh-submono}
\end{equation}%
where $K$ is a constant that comes from the geometry of the deposited structure and $\varepsilon_i^2$ is measured through imaging techniques, as discussed above. In our case, it is a good approximation to consider those layers as 2d hexagonal close packed structure, and thus $K=\frac{2\pi}{3\sqrt{3}}\cdot R$, where $2R$ is the diameter of the colloidal particles. Here, e.g. in Figure~\ref{deposits_analysis}B we can see monolayer and bilayer deposits in different colors, therefore we measure total CEH as: 
  
\begin{equation}
\mbox{CEH}_{total}\approx \mbox{CEH}_{1} + \mbox{CEH}_{2}
\label{CEHtotal}
\end{equation}
where $\mbox{CEH}_{i}$ is the thickness of a continuous homogeneous layer whose volume is the same as the particles deposited on layer $i$ (Figure~\ref{cehsketch}).

\subsection{Magnetorheology by spin-coating.}

Pichumani {\it et al.}\cite{Pichumani2013} used colloids model and compared two experiments results obtained in the same conditions, except for the applied magnetic fields. That comparison resulted in the ratio of CEH, measured from the experiments performed with and without magnetic fields, using eq.~\ref{eqcregan-ceh}.

The kinematic viscosity $\nu$ is the only parameter which may depend on the applied magnetic field, and we can take the ratio of eq.~\ref{eqcregan-ceh} when a field is applied over the same at zero-field. That gives information of the relative effective viscosity:
\begin{equation}
\frac{\nu(H)}{\nu(H=0)}=\left[\frac{\mbox{CEH}(H)}{\mbox{CEH}(H=0)}\right]^3.
\label{mag-gen4}
\end{equation} 

\section{Experimental results and discussion}
\begin{figure}[!ht]
\centering
\includegraphics[width=8cm]{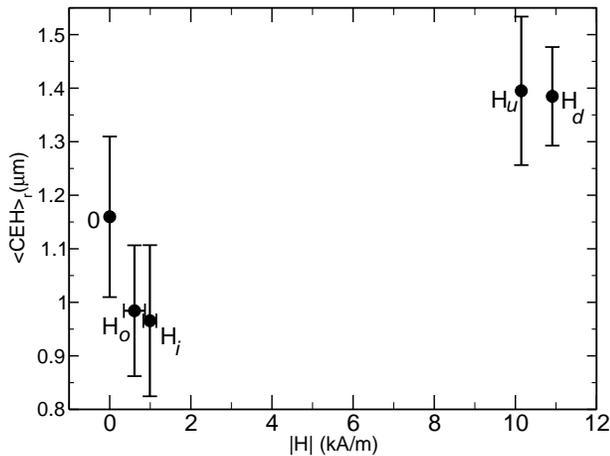}
\caption{Mean compact equivalent height (CEH) obtained from spin-coated deposits of superparamagnetic particles suspension in MEK of concentration of 12.82\% (v/v) (diameters of silica coated magnetite particles are 1.51~$\mu$m) at 2000~rpm spinning, without and with different applied magnetic field configurations.}
\label{CEH}
\end{figure}

Figure~\ref{deposits_analysis}C--D are typical images, for a given sample, that allow to measure $\varepsilon_i^2$, of the dried spin-coated deposits at different distances $r$ from the center of rotation. The particulate nature of the solute requires to translate the raw measured occupation factors into compact equivalent heights CEHs (using eqs.\ \ref{ceh-submono}, \ref{CEHtotal}), and their results are shown in Figures~\ref{cehvsrmm} and~\ref{CEH}. We obtain Figure~\ref{CEH} by averaging CEH of Figure~\ref{cehvsrmm} over $r$ for without and with magnetic field of different configurations (detailed procedure is explained elsewhere \cite{Aslam2016_thesis}). In both figures, we see that the final thickness and the uniformity of deposits depend on the distance from the center of rotation \cite{Giuliani2010} as well as on the configuration of applied magnetic fields. For the uniform applied magnetic fields (H$_{u}$ and H$_{d}$) we have larger values of CEH than for the non-uniform magnetic fields (H$_{o}$ and H$_{i}$).

\begin{figure}[!ht]
  \centering
  \includegraphics[width=8cm]{fig5}
  \caption{Compact equivalent height (CEH) as a function of increasing radial distance from the center of spinning. Squares and circles  represent results from the experiments performed at two different applied magnetic field configurations (H$_{u}$ and H$_{o}$) respectively and filled black circles for zero-field. Spinning rate for all cases is 2000~rpm.}
\label{cehvsrmm}
\end{figure}

\begin{figure}[!ht]
  \centering
  \includegraphics[width=8cm]{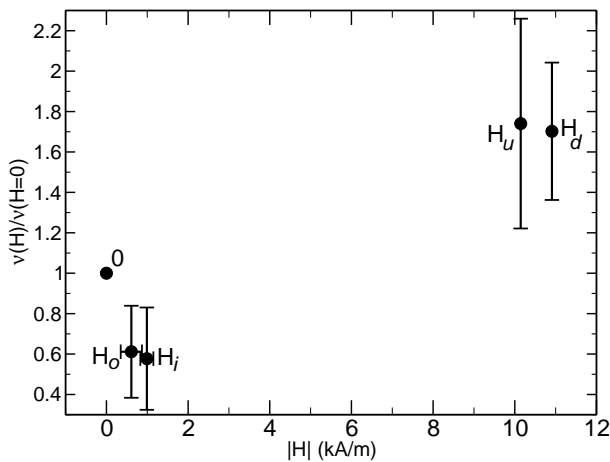}
  \caption{Normalized efective magnetoviscosity as a function of different magnetic field configurations. The concentration of the suspension is 12.82\% (v/v), and the diameters of superparamagnetic particles are 1.51~$\mu$m. The spinning rate is 2000~rpm.}
  \label{viscosityvsH}
\end{figure}

From eq.~\ref{mag-gen4} and Figure~\ref{CEH}, we obtain Figure~\ref{viscosityvsH}. On the one hand, we can observe a large increase in the magnetoviscosity for uniform magnetic fields as compared to the viscosity at zero-field. On the other hand, for non-uniform magnetic fields a decrease in the effective viscosity is observed, also compared to the viscosity at zero-field. In these configurations (H$_{i}$ and H$_{o}$) there is a gradient of the magnitude of the magnetic field pointing outwards in the radial direction. As a consequence, there is a magnetic force on each dipole (superparamagnetic particle) also in that direction. This is part of the magnetophoretic effect, that leaves less deposits on the rotating substrate (see H$_{o}$ curve in figure \ref{cehvsrmm}). This can be regarded as a result of a smaller effective viscosity, which actually is not related to the true magnetoviscosity of the system.

\section{Summary and concluding remarks}
We report results on the spin-coated deposits of a moderately concentrated superparamagnetic colloid in volatile solvents when different magnetic fields are applied. They are compared with those obtained from the experiment performed at zero-field.

The case of uniform (homogeneous) magnetic fields (H$_{u}$ and H$_{d}$) shows that the magnetorheological spin-coating technique \cite{Pichumani2013} is consistent, and also useful for moderately concentrated superparamagnetic colloids. It is possible then, to measure the CEH statically, and compare it to the zero-field CEH to obtain the magnetoviscosity.

The case of non-uniform magnetic field configurations (H$_{i}$ and H$_{o}$) shows a decrease in the effective magnetoviscosity when calculated through the colloidal deposits (CEH). This happens because of the gradient of magnetic field that induces magnetophoresis effect and, consequently, enhances the hydrodynamical flows out of spinning center. Thus, under non-uniform magnetic fields the colloids flow as they were less viscous, although the (true) magnetoviscosity is higher than the viscosity at zero-field.

Furthermore, our results indicate that the two uniform (H$_{u}$ versus H$_{d}$) and the non-uniform (H$_{i}$ versus H$_{o}$) magnetic field configurations lead to symmetric results in the deposits, within the experimental error, as shown in Figures~\ref{CEH} and~\ref{viscosityvsH}. 

Although we have presented the main effect of external inhomogeneous magnetic fields on the effective magnetoviscosity of a magnetic colloid in spin-coating, further research is required to identify the dynamics of the drying suspension. Moreover, with the methods discussed here more experiments have to be done to study effective magnetoviscosity in more concentrated suspensions at applied non-uniform magnetic, which might be important for the technological applications of superparamagnetic colloids.

\begin{acknowledgments}
This work was supported by the Spanish AEI (Grant n.\ FIS2014-54101-P).
\end{acknowledgments}

\bibliography{scHsilicadiffconf}

\end{document}